\documentclass[12pt]{article}
\usepackage[pctex32]{graphics}
\begin{document}
\begin{center} {\Large \bf   Hawking Radiation of a Quantum Black Hole in an Inflationary Universe}\\
\vspace{0.5cm}
                     Wung-Hong Huang\\
                       Department of Physics\\
                       National Cheng Kung University\\
                       Tainan,Taiwan\\
\end{center}
\vspace{0.5cm}
\begin{center} {\Large \bf  Abstract} \end{center}
The quantum stress-energy tensor of a massless scalar field propagating in the two-dimensional Vaidya-de Sitter metric, which describes a classical model spacetime for a dynamical evaporating black hole in an inflationary universe, is analyzed. We present a possible way to obtain the Hawking radiation terms for the model with arbitrary functions of mass.  It is used to see how the expansion of universe will affect the dynamical process of black hole evaporation. The results show that the cosmological inflation has an inclination to depress the black hole evaporation. However, if the cosmological constant is sufficiently large then the back-reaction effect has the inclination to increase the black hole evaporation. We also present a simple method to show that it will always produce a divergent flux of outgoing radiation along the Cauchy horizon where the curvature is a finite value. This means that the Hawking radiation will be very large in there and shall modify the classical spacetime drastically. Therefore the black hole evaporation cannot be discussed self-consistently on the classical Vaidya-type spacetime. Our method can also be applied to analyze the quantum 
stress-energy tensor in the more general Vaidya-type spacetimes.
\vspace{0.5cm}
\begin{flushleft}
E-mail:  whhwung@mail.ncku.edu.tw\\
Class. Quantum Grav. 9 (1992) 1199-1209\\ 
--------------------------\\
{\bf Proper boundary will lead to anti-evaporation of schwarzschild-de Sitter black holes, as corrected in Class. Quantum Grav. 11 (1994) 283.}
\end{flushleft}
\newpage
\section  {Introduction}
Hawking [1] discovered that a black hole formed by collapsing matter will 
emit particles like a black body with temperature proportional to its surface gravity.  As the original calculation is done for a static black hole, which is valid only in the case of the small-evaporation limit, we shall, for self-consistency, take account of the back reaction by solving the semiclassical Einstein equation [2]
$$ R_{\mu\nu} -{1\over2} g_{\mu\nu} R = 8\pi <T_{\mu\nu}> ,\eqno{(1.1)}$$
where $<T_{\mu\nu}>$ is the quantum stress-energy tensor which is a regularized one evaluated in a suitable vacuum [3-5].  As  $<T_{\mu\nu}>$ is a geometrical object which depends on the geometry of the spacetime, equation (1.1) becomes a highly non-linear set of  coupled partial differential equations and solving it is a very difficult task even after several approximations have been adopted [6]. In view of this, as a second-best method, many authors have investigated the black-hole evaporation on the classical spacetime background which represents, in any case, a dynamically evaporating black hole [7-12]. As was found many years ago, the Vaidya metric [13] and Vaidya-Bonnor metric [14] are the most suitable spacetimes describing an evaporating neutral and charged black hole, respectively.

Because the black hole really is a cosmological object, it is worthwhile to examine
the effect of cosmological evolution on the process of black-hole evaporation [15-17].  In [15], Davies et. al. investigated the thermodynamics of a black hole in 
the Reissner-Nordstrom-de Sitter space.  In [16] Balbinot et. al. discussed the cosmological boundary condition for a black hole.  In [17] Mallett constructed the Vaidya-de Sitter metric
$$ds^2=-\left( 1- {2M(v)\over r} - {\Lambda\over 3} r^2\right) \, dv^2 + 2 dv\, dr + r^2(d\theta^2 + sin^2\theta)\, d\varphi^2 ,\,\eqno{(1.2)}$$
where $M$ is a function of the Eddington-Finkelstein-type advanced time $v$.  After a straightforward tensor calculation one sees that this metric will satisfy the Einstein equation with a cosmological constant $\Lambda$  and source term which represents the matter streaming radiating inwards; thus, the Vaidya-de Sitter metric could be used to model the classical spacetime of an evaporating black hole in a de Sitter universe. Mallett uses this metric to analyze the process of  black-hole evaporation and finds that the hole evaporation will be depressed by the  cosmological inflation.

In this paper we will analyze the quantum stress-energy tensor of a massless 
scalar with an arbitrary function of $M(v)$. Our goal is to see how the universe 
expanding will affect the process of black-hole evaporation. Before so doing, one 
should notice that in the Vaidya spacetime, because of the technical difficulties, we have as yet no ability to exactly evaluate  $<T_{\mu\nu}>$ except for the linear model [9] in which the mass $M$ is a linear function of the advanced time. For the model in the Vaidya-de Sitter spacetime, the situation becomes worse, because even for the linear model we cannot obtain the exact formula of  $<T_{\mu\nu}>$. This is unfortunate, since knowing the exact function form of  $<T_{\mu\nu}>$ would let us see the back-reaction effects on black-hole evaporation immediately.  

The present work does not try to evaluate the components of  $<T_{\mu\nu}>$  exactly.  Rather, we will present a simple method to find the Hawking radiation terms among the quantum stress-energy tensor of a massless scalar field propagating in the Vaidya-de Sitter spacetime for the general case with arbitrary functions of  $M(v)$.  As we know, the Hawking radiation term is a component of the quantum stress-energy tensor, which represents the outgoing energy flow and has a finite value at infinity.  From the obtained Hawking radiation terms we can then give a discussion about the back  reaction effect on an evaporating black hole immersed in an inflationary universe. Our results show that the cosmological inflation has an inclination to depress the black hole evaporation as claimed in [17]. However, we find the effect of back reaction, i.e., the mass variation, can increase the black-hole evaporation if the cosmological constant is sufficiently large.  We also present a simple method to show that it will always produce a divergent flux of outgoing radiation along the Cauchy horizon. This means that the Hawking radiation will be very large there and will modify the classical spacetime drastically. Therefore the black-hole evaporation cannot be discussed self-consistently on the classical Vaidya-type spacetime.

This paper is organized as follows. In section 2, the model is set up and formula for calculating the quantum stress-energy tensor are constructed. In section 3, we present a simple method to find the Hawking radiation terms among the $<T_{\mu\nu}>$  and then use the results to discuss the effect of universe expansion on the process of black-hole evaporation. The universal behaviour of producing a divergent flux of outgoing radiation along the Cauchy horizon in the Vaidya spacetime is proved in section 4.  Section 5 is devoted to discussion.

\section  {Model and formalism}
We will calculate the quantum stress-energy tensor for a massless scalar field propagating in a two-dimensional spacetime which is that obtained by taking a 
constant $\theta$ and a constant $\varphi$ in the four-dimensional spacetime described by the Vaidya-de Sitter metric in equation (12). This is because a two-dimensional spacetime is conformally flat and a general method has been constructed to obtain the renormalized stress-energy tensor of a massless scalar field [3, 18]. 

To proceed, we first set up the double-null coordinates in the three regions.\\
{\it Region I}
$$ds_1^2= - (1 -{1\over3}\Lambda r^2) dv^2+ 2 dv dr = - (1 -{1\over3}\Lambda r^2)du_1 dv, ~~~~~~~~~~v < 0, \eqno{(2.1)}$$
where we have defined 
$$du_1 = dv - (1 -{1\over3}\Lambda r^2)^{-1},\eqno{(2.2)}$$
and thus
$$u_1= v-2 \sqrt {3/\Lambda}\,\,coth^{-1}\sqrt {3/\Lambda} \,r, ~~~~ if ~~ \sqrt {3/\Lambda}\,\, r >1 \eqno{(2.3a)}$$
$$u_1= v-2 \sqrt {3/\Lambda}\,\,tanh^{-1}\sqrt {3/\Lambda} \,r, ~~~~ if ~~ \sqrt {3/\Lambda} \,\,r <1. \eqno{(2.3b)}$$
\\
{\it Region II}
$$ds_2^2 = - (1 - {2M(v)\over r}-{1\over3}\Lambda r^2) dv^2+ 2 dv dr ~~~~~~~~~~~~~~~~~~~~~~~~~~~~~~~~~~~~$$
$$ =  - (1 - {2M(v)\over r}-{1\over3}\Lambda r^2)\, G^{-1}(v,r) du_2 \,dv \equiv - D(v,r) du_2 ,\,dv   \,\eqno{(2.4)}$$
where we have defined
$$du_2 = G(v,r) dv - 2G(v,r) \left(1 - {2M(v)\over r}-{1\over3}\Lambda r^2\right)^{-1} dr\,\eqno{(2.5)}$$
and thus G is an integrating factor which satisfies
$${\partial G \over \partial r} + 2 {\partial\over\partial r}  \left[G(v,r) \left(1 - {2M(v)\over r}-{1\over3}\Lambda r^2\right)^{-1}\right] = 0.\eqno{(2.6)}$$
\\
{it Region III}
$$ds_1^2= - (1 -{1\over3}\Lambda r^2) dv^2+ 2 dv dr = - (1 -{1\over3}\Lambda r^2)du_1 dv, ~~~~~~ v > 0, \eqno{(2.7)}$$
where we have defined 
$$du_1 = dv - (1 -{1\over3}\Lambda r^2)^{-1},\eqno{(2.8)}$$
and thus
$$u_1= v-2 \sqrt {3/\Lambda}\,\,coth^{-1}\sqrt {3/\Lambda}\, r, ~~~~ if ~~ \sqrt {3/\Lambda}\,\, r >1 \eqno{(2.9a)}$$
$$u_1= v-2 \sqrt {3/\Lambda}\,\,tanh^{-1}\sqrt {3/\Lambda} \,r, ~~~~ if ~~ \sqrt {3/\Lambda} \,\,r <1. \eqno{(2.9b)}$$
The above model is initially ($v < 0$) in de Sitter spacetime. Then, at $v = 0$ an imploding $\delta$-functional shell of null fluid with positive mass $M(0) = m_0$ forms a black hole [9]. Next, during the interval $0 < v <v_0$, negative-energy-density null fluid falls into the hole to evaporate it gradually. The evaporation rate depends on $M(v)$ which is assumed to be an arbitrarily continuous function. At $v =v_0$, the black hole completely vanishes and the final geometry ($v >v_0$) is again in de Sitter spacetime. A Penrose diagram for such a model can be depicted  by piling up  sliced Schwanschild-de Sitter spacetime (with various mass) so that  is continuous. Note that the Penrose diagram of de Sitter and Schwanschild-de Sitter spacetimes can be found in the paper of Gibbons and Hawking [19]. (See also [20].)

Note that relation (2.3a) is adopted to evaluate the Hawking radiation term (which is a component of the stress tensor at $r\rightarrow \infty$) in region 11, as the limiting function $G(v, r\rightarrow \infty)$ is a concern. On the other hand, the relations (2.3b) and (2.9b) are adopted to analyse the divergent behaviour along the Cauchy horizon in region Ill, as the limiting function $G(v, r\rightarrow 0)$ is the concerning one now. (See section 4.)

To determine the stress-energy tensor $<T_{\mu\nu}>$  in region I1 we need a relation between $u_1$ , and $u_2$. This can be found from the match condition [9]. Matching the coordinate across $v =0$ gives the following differential equation 
$${du_1\over du_2} = - E(u_1)\, sinh^2\left(\sqrt{\Lambda\over3}{u_1\over 2}\right),$$
$$E(u_1) = D\left(0,- \sqrt{3\over \Lambda}\, coth\left(\sqrt{\Lambda\over3}{u_1\over 2}\right)\right). \eqno{(2.10)}$$
Thus
$$ds_2^2 = - D(v,r) E^{-1}(u_1) \,cosech^2\left(\sqrt{\Lambda\over3}{u_1\over 2}\right)\,du_1\,dv . \eqno{(2.11)}$$

The two-dimensional stress-energy tensor for a quantized massless scalar field
could now be evaluated by relating the null coordinates to a suitable set $(\tilde v, \tilde u)$  in which the vacuum state is defined [3,4]. In our model, the scalar fields modes for  the vacuum have the form $exp(-iwv)$ in the infinite past.  However, the metric in equation (2.1) shows that our spacetime will become a de Sitter type and not flat in the asymptotical past. This means that we cannot evaluate the quantum stress tensor in the well known 'Unruh vacuum' [3] which reduces to the Minkowski space in the past. The stress tensor found below, therefore, shall be regarded as that evaluated in the  background of the de Sitter universe, i.e. we choose $(\tilde v, \tilde u)$  = $(v, u_1)$, and our results do not contain that of the particle created in the de Sitter spacetime [19,20].  Keeping this meaning in mind we then, after the typical procedure [9], obtain the renormalized stress-energy tensor of a massless scalar field
$$ <T_{\mu_2\nu_2}> = {1\over 24\pi}\left({1\over4}DD_{,rr}-{1\over8}(D_
{,r})^2 + sinh^4\left(\sqrt{\Lambda\over3}{u_1\over 2}\right)  \right.$$
$$\times \left. \left({1\over2}(E_{,u_1})^2 - E E_{,u_1u_1} - \sqrt{\Lambda\over3}\,E E_{,u_1}\, coth\left(\sqrt{\Lambda\over3}{u_1\over 2}\right)   \right) - {\Lambda\over6}E^2\right) \eqno{(2.12)}$$

In the same way, to determine the stress-energy tensor  $<T_{\mu_3\nu_3}>$ in region I11 we need the relations between $u_1, u2$ and $u_3$. These can be found from the match  conditions.   The coordinates match across $v = 0$ and $v = v_0$ giving the following differential equations
$${du_1\over du_2} = - F_1(u_1)\, cosh^2\left(\sqrt{\Lambda\over3}{u_1\over 2}\right),$$
$$F_1(u_1) = D\left(0,- \sqrt{3\over \Lambda}\, tanh\left(\sqrt{\Lambda\over3}{u_1\over 2}\right)\right). \eqno{(2.13)}$$
$${du_3\over du_2} = - F_2(u_3)\, cosh^2\left(\sqrt{\Lambda\over3}{u_1\over 2}\right),$$
$$F_2(u_3) = D\left(0,- \sqrt{3\over \Lambda}\, tanh\left(\sqrt{\Lambda\over3}{u_3-v_0\over 2}\right)\right). \eqno{(2.14)}$$
Thus
$$ds_2^2 = - \left(1- {\Lambda\over3}\, r^2 \right) F_2 (u_3) F^{-1}(u_1)\,cosh^2\left(\sqrt{\Lambda\over3}{u_3- v_0\over2}\right) du_1dv \equiv - \left(1- {\Lambda\over3}\, r^2 \right) W\,du_1dv. \eqno{(2.15)}$$

Using the same definition of the vacuum state discussed before we then, after the typical procedure [9 ], obtain the renormalized stress-energy tensor of a massless scalar field
$$<T_{\mu_3\nu_3}> = {1\over 24\pi}W^{-2} \left[Z_{,u_1}- {1\over2}Z^2\right] , ~~~~~ Z \equiv \left[ln \left(1- {\Lambda\over3}\, r^2 \right)W\right]_{,u_1}. \eqno{(2.16)}$$
Note that if $M=0$ then,as can be easily found $<T_{\tilde u \tilde u}>_0 = <T_{v v}>_0 = - \Lambda/144\pi $  and $<T_{\tilde u v}>_0 = <T_{v \tilde u}>_0 = 0$. Neglecting the constant value of  $<T_{\mu \nu}>_0$, then, in the region I  all components  of $<T_{\mu \nu}>$ are zero; in region III the only non-zero component of $<T_{\mu \nu}>$ is  that expressed in equation (2.12); in region II although all components of $<T_{\mu \nu}>$ are non-vanishing, only $<T_{u_2  u_2}>$ will give a non-zero value as $r\rightarrow \infty$ , This is the Hawking radiation term which represents the outgoing energy flow at infinity.

In the following section we will present a possible way to obtain the Hawking
radiation terms in for the model with arbitrary function of mass $M(v)$. From
the result the effects of the mass variation on the evaporation of a black hole immersed in de Sitter spacetime are discussed. Then, in section 4 we will show that $<T_{u_2  u_2}>$ will always produce a divergent flux of outgoing radiation along the Cauchy  horizon.

\section  {Hawking radiation terms in the stress tensor}
From equation (2.12) it is seen that to evaluate $<T_{u_2  u_2}>$ we need to know the function of $D(v, r)$, or $G(v, r)$. It is unfortunate that the function $G(v,r)$ which is the solution of equation (2.6) could not be found even for a simple choice of the linear functions $M(v)$. However, as we are only interested in the Hawking radiation terms we do not need to know so much. From the formula of $<T_{u_2  u_2}>$ expressed in equation (2.12) we see that, to evaluate the first two terms, only a knowledge of the limiting function $G(v, r\rightarrow\infty)$ is necessary, while the last four terms can be evaluated with  merely the limiting function $G(v, r\rightarrow 0)$. We will show below how to find these two limiting functions.

\subsection { Radiation in initial stage}
We first solve equation (2.6) at initial time $v\rightarrow 0$ to find the function $G(v\rightarrow 0,r)$. Substituting the expansions
$$M(v) \approx  m_0 + m_1 v + m_2 v^2, ~~~~~~\eqno{(3.1a)}$$
$$G(v,r) \approx  g_0(r) + g_1(r) v + g_2(r) v^2, \eqno{(3.1b)}$$
into equation (2.6) we obtain the equation
$$\dot g_0 + 2 \left(1-{2m_0\over r}- {\Lambda\over3}\, r^2 \right)^{-1} g_1 + {4m_1\over r} \left(1-{2m_0\over r}- {\Lambda\over3}\, r^2 \right)^{-2} g_0 = 0 , \eqno{(3.2)}$$
$$\dot g_1 + 4 \left(1-{2m_0\over r}- {\Lambda\over3}\, r^2 \right)^{-1} g_2 + {8m_1\over r} \left(1-{2m_0\over r}- {\Lambda\over3}\, r^2 \right)^{-2} g_1 ~~~~~~~~~~~~~~~~~$$

$$ + {16m_1^2\over r} \left(1-{2m_0\over r}- {\Lambda\over3}\, r^2 \right)^{-3} g_0 + {8m_2\over r} \left(1-{2m_0\over r}- {\Lambda\over3}\, r^2 \right)^{-2} g_0 = 0 . \eqno{(3.3)}$$
Choosing $g_0(r) = 1$ we then obtain
$$g_1(r) = -{2m_1\over r} \left(1-{2m_0\over r}- {\Lambda\over3}\, r^2 \right)^{-1}.\hspace{5cm} \eqno{(3.4a)}$$
$$g_2(r) = -{m_1\over2r^2}-  \left({m_0 m_1\over r^3}+{2m_2\over r}- {\Lambda\over3} m_1 \right) \left(1-{2m_0\over r}- {\Lambda\over3}\, r^2 \right)^{-1}. \eqno{(3.4b)}$$
Thus we obtain
$$D(v\rightarrow 0,r) =  \left(1-{2m_0\over r}- {\Lambda\over3}\, r^2 \right) +\left({1\over 2r^2}- {\Lambda\over2}\right) \, m_1 \, v^2 . \eqno{(3.5)}$$
(Note that the choice of $g_0(r) = 1$, which implies the above function of $D(v\rightarrow 0,r)$, must be consistent with the function $D(v, r\rightarrow 0)$solved in section 3.2.)  

Substituting the above expression into equation (2.12) we then obtain the Hawking radiation terms in the initial stage
$$<T_{u_2  u_2}> \rightarrow {1\over 24\pi}(H_0,\tilde H_b),  \eqno{(3.6)}$$
where
$$H_0 = {\Lambda^2\over 9} \,m_0^2  \,sinh^4 \sqrt{\Lambda\over3}{u_1\over2}\left[2\left(1-sech^2\sqrt{\Lambda\over3}{u_1\over2} \right)^2 + sech^4\sqrt{\Lambda\over3}{u_1\over2}\right] ~~~$$
$$+{\Lambda\over3} \, m_0 \left[tanh \sqrt{\Lambda\over3}{u_1\over2} \left(2 +  sech^2\sqrt{\Lambda\over3}{u_1\over2} + 2 sinh^2\sqrt{\Lambda\over3}{u_1\over2}\right) \right]\eqno{(3.7)}$$
$$\tilde H_b= {\Lambda^2\,m_1\over12}\,v^2 - {\Lambda\over6} \eqno{(3.8)}$$
Note that as $u_1 <0$ the value of $H_0$ is negative.  Also, according to the known result, $\dot M(v) \sim - M^{-2}(v)$ at the initial stage, thus $m_1 < 0$ and the value of $\tilde H_b$ is negative too.  We therefore see that the terms coupling the cosmological constant with the mass or mass variation are always negative. These mean that the cosmological inflation has an inclination to depress the black hole evaporation. The conclusions are consistent with [17].

\subsection {Radiation at any time}
To investigate the black hole evaporation at any time in region II we shall find the limiting function $G(v, r \rightarrow \infty)$ which satisfies equation (2.6). Mathematically, equation (2.6) is a first-order partial differential equation and its solution can be found with the help of the character equation
$$dv =  \left(1-{2m_0\over r}- {\Lambda\over3}\, r^2 \right) {dv\over 2} = {-r\over 4\, dM /dv} \left(1-{2m_0\over r}- {\Lambda\over3}\, r^2 \right)\,d \,lnG. \eqno{(3.9)}$$
When $r \rightarrow \infty$, the first equation gives a simple solution
$${6\over \Lambda r} - v \approx c_1 , \eqno{(3.10)}$$
where $c_1$ is an integration constant. With this relation the second equation becomes (the approximation $r \rightarrow \infty$ has been taken)
$$d\, ln\, G^{-1} = {-\dot M \, \Lambda^2 \over 36} \, (v + c_1)^3\,dv. \eqno{(3.11)}$$
The above equation can now be integrated formally. Again, replacing the $c_1$
by the relation equation (3.10) and taking the limiting $r \rightarrow \infty$ we finally obtain
$$G^{-1}(v,r) \approx  c_2\, e ^{\Lambda^2 A/36}\left[1 + {\Lambda \over 2 r}B + {1\over r}\left({\Lambda^2 \over 8}B^2 - C\right) \right], \eqno{(3.12)}$$
where $A$, $B$ and $C$ are defined by
Note that although the integration constant c2 may be an arbitrary one it 
$$ A \equiv - \int_0^v\, \dot M(\tilde v) (v- \tilde v)^3 \, d\,\tilde v\eqno{(3.13a)}$$
$$ B \equiv - \int_0^v\, \dot M(\tilde v) (v- \tilde v)^2 \, d\,\tilde v\eqno{(3.13b)}$$
$$ C \equiv - 3 \int_0^v\, \dot M(\tilde v) (v- \tilde v) \, d\,\tilde v .\eqno{(3.13c)}$$
Note that although the integration constant $c_2$ may be an arbitrary one it must be chosen to be consistent with the function $G(v\rightarrow 0,r)$ found in section 3.1. Therefore $c_2$ shall be chosen as one.

Using the above obtained limiting functions $G(v, r\rightarrow \infty)$ and $G(v\rightarrow 0,r)$ we finally from equation (2.12) obtain the Hawking radiation terms 
$$<T_{u_2 u_2}> \rightarrow {1\over 24\pi}(H_0 + H_b), ~~~~~as ~~r \rightarrow \infty.\eqno{(3.14)}$$
where $H_0$ being defined in equation (3.7) does not depend on the advance time $v$, and  $H_b$ showing the effect of back reaction is defined by
$$H_b \equiv \left({\Lambda^4\over 288} B^2-{\Lambda^2\over 18}C -{\Lambda\over 6}\right)\, e^{\Lambda^2 A/18}.\eqno{(3.15)}$$
with A, B and C defined in equation (3.13).

From equation (3.14) we see that when the cosmological constant $\Lambda$ is small then $H_b= -\Lambda /6- \Lambda^2C/18$ . Because $C$ is positive (see equation (3.13b)) the back-reaction effect will depress the black-hole evaporation. (If both $\Lambda$ and $v\rightarrow 0$ then the above results give the value of $<T_{u_2 u_2}>$ as described by equation (3.6).) Combining the fact that $H_0$ defined in equation (3.7) is negative, we thus see that the terms coupling the cosmological constant with the mass or mass variation are always negative. This means that the cosmological inflation has an inclination to depress the black hole evaporation, as claimed in [17]. On the other hand, i f the  cosmological constant $\Lambda$ becomes very large then it is apparent that $H_b$, will become a positive function. This means that a time-dependent mass term entering the Vaidya-de Sitter metric will increase the Hawking radiation if the cosmological constant is sufficiently large, as claimed in the introduction. (This conclusion does not agree with [17].) Notice that $H_0$, being always
positive, is a function of $\Lambda$, $M(0)$ and $u_1$; while $H_b$, being positive if $A$ is very large, is a function of  $\Lambda$, $\dot M(v)$ and $v$.

Finally we shall make two remarks. (i) One may wonder why the property of
increasing the black-hole evaporation by the back-reaction effect when the cosmological constant is large does not show in section 3.1. The reason is that the expansion at $v= 0$, i.e., in initial stage, will correspond to the expansion about $\Lambda = 0$, according to the formulation in section 3.1. This may be seen from the fact that taking more expansion terms in equation (3.1) will let the function of $D(v \rightarrow 0. r)$ in equation (3.5) contain the terms proportional to $\lambda$,... etc. (ii ) Once $\lambda = 0$ was taken, then nothing would be obtained. One may wonder why the result  [12] for the model in the Vaidya spacetime will not be obtained in this limit. The reason is that we adopt equation (2.3a) rather than equation (2.3b) during the formulation, as we are interested in the Hawking radiation term which is defined at infinity, then equation (2.3a) does not define itself at $\lambda = 0$.

\section  {Divergence of the stress tensor along the Cauchy horizon}
The analyses The analyses for the linear model [9] in the Vaidya or Vaidya-Bonnor spacetime have found that the stress-energy tensor in region III will be divergent as the Cauchy horizon is approached, i.e., $u_3\rightarrow v_0$.  This divergent property has been shown to be very general.   In this section we present a simple way to prove that the quadratic divergence of the quantum stress-energy is universal for the model described in the Vaidya-de Sitter spacetime.

From equation (2.16) we see that, as we are only interested in the stress-energy
tensor as the Cauchy horizon is approached, i.e., $u_3\rightarrow v_0$, the only function to be found is the limiting form of $F_2(u_3 \rightarrow  u_o) = D(v_0, -{1\over2}(u_3 - v_0) \rightarrow 0)$ and its corresponding limiting
function of $F_l(u_l)$. We can find these two functions easily as shown below.

In the limit of  $r \rightarrow  0$ equation (2.6) becomes
$${\partial\, G\over \partial\, r} \approx r \,{\partial\,\over \partial\, v}\, (G/M), ~~~~~~~ r \rightarrow 0, \eqno{(4.1)}$$
and a simple solution we have is
$$G(v_0, r \rightarrow 0) \approx c(1+r^2). \eqno{(4.2)}$$
where $c$ is an integration constant. Choosing a suitable value of $c$ we have the solution
$$F_2(u_3 \rightarrow v_0) = D(v_0, -{1\over2}(u_3-v_0)\rightarrow 0 ) \approx{1\over u_3-v_0}. \eqno{(4.3)}$$
Also, using equations (2.13) and (2.14) we have the relation
$$\int ^{u_1}F_1(\tilde u_1) \,cosh^2\left(\sqrt{\Lambda\over3}{\tilde u_1\over 2}  \right) \, d \tilde u_1= \int ^{u_1}F_2^{-1}(\tilde u_3) \,sech^2\left(\sqrt{\Lambda\over3}{\tilde u_3-v_0\over 2}  \right) d \tilde u_3$$
$$~~~~~~~~~~~~~~~~~~~\approx  \int ^{u_1} (\tilde u_3-v_0)\, d \tilde u_3, ~~~~ as ~~u_3 \rightarrow v_0 , \eqno{(4.4)}$$
which tells us that as $u_3 \rightarrow v_0$  the value of $u_1$ approaches a value (say) so $s_0$ and the integrated function is finite. Thus we have the approximation
$$F_1( u_1) \,cosh^2\left(\sqrt{\Lambda\over3}{\tilde u_1\over 2}  \right)\approx a + b (u_1 - s_0)^\alpha,  ~~~~~~ \alpha > -1 , ~~~~ as ~~u_1 \rightarrow s_0, \eqno{(4.5)}$$

Substituting the above obtained limit functions $F_2(u_3 \rightarrow v_0)$ and $F_1(u_2 \rightarrow  s_0)$ into equation (2.16) and through some careful analyses we finally obtain the results
$$<T_{u_3 u_3}> \rightarrow {1\over 16\pi (u_3-v_0)^2}, ~~~~~~~~~~as ~~u_3 \rightarrow v_0, \eqno{(4.6)}$$
The above relation shows that the divergence along the Cauchy horizon is a 
universal property. However, a divergence appearing at the Cauchy horizon where the curvature has a finite value seems physically serious. In fact, this tells us that the Hawking effect will be vary large there and will modify the classical spacetime drastically. Therefore, the evaporation of a quantum black hole cannot be discussed self-consistently on the classical Vaidya-de Sitter spacetime.

\section  {Conclusion}
Several years ago, Davies et al [l5] investigated the thermodynamics of a black hole in the Reissner-Nordstrom-de Sitter space. They used the black-hole temperature and cosmological horizon temperature to discuss how the thermal radiation will flow between the hole and universe. However, as is well known, a rigorous description of black-hole evaporation requires the hack-reaction effect to be seriously taken into account. This then leads Mallett [17] to construct the Vaidya-de Sitter metric to model the classical spacetime of an evaporating black hole in a de Sitter universe. Mallett uses this metric to analyse the dynamical behaviour of a black hole immersed in an inflationary universe and finds that the process of black-hole evaporation will be depressed by the cosmological inflation.

An apparent way to analyze the back-reaction effect on an evaporating black hole
immersed in an inflationary universe is to analyze the quantum stress-energy tensor of a massless scalar field propagating in the two-dimensional Vaidya-de Sitter metric. At first sight, because, to our knowledge, nobody has the ability to solve equation (2.6) except for the step model in which $M (v)$ is a constant, it seems that such a way is hopeless. In this paper, we do not try to solve equation (2.6); rather, we have succeeded in presenting a simple method to find the Hawking radiation term, which is a component of the quantum stress-energy tensor and represents the outgoing energy flow at infinity. From the obtained Hawking radiation terms we then give a discussion about the back-reaction effect on an evaporating black hole immersed in an inflationary universe. Our results show that the cosmological inflation has an inclination to depress the black hole evaporation as claimed in [17]. However, we find that the effect of back-reaction,
i.e., the mass variation, can increase the black-hole evaporation if the cosmological constant is sufficiently large. We also have presented a simple method to show that it will always produce a divergent flux of outgoing radiation along the Cauchy horizon where the curvature is a finite value. This means that the Hawking radiation will be very large there and will modify the classical spacetime drastically. Therefore we conclude that the black-hole evaporation cannot be discussed self-consistently on the classical Vaidya-type spacetime.

Finally we want to make a remark. As the prescription presented in this paper is
very general and simple, it will be (at least, we hope) helpful to analyze the dynamical behavior of a black hole in more general, more complex and relatistic Vaidya-type spacetimes.

More references can be found in [22-24].


\newpage
\begin{center} {\large \bf  References} \end{center}
\begin{enumerate}
\item  Hawking S W 1975 Commun. Marh. Phys. 43 199.
\item  Birrcll N D and Dabies PC W 1982 Quonlum Fieldr in CuruedSpoee (Cambridge: Cambridge University press). 
\item  Unruh W G 1976 Phys. Rev. D 14 870;\\
          Davies P C W, Fulling S A and Unruh W C 1916 Phys. Rev. D 13 2720;\\
\item  Boulware D G 1975 Phys. Rev, D I 1 1410; D 12 350;\\
          Harile J B and Hawking S W 1976 Ph.vs. Rea D 13 2188;;
          Israel W 1976 Phys. Lnr. 51A 107.
\item  Candelas P 1980 Phy. Rev. D 21 2185.
\item  York J 1985 Phy. Rev. D 31 775.
\item  Bardeen J M 1981 Phyr Rev. Len. 46 382.
\item Sullivan B T and Isreel W 1980 Phys. Lett. 19A 371.
\item Hiscock W A 1981 Phys. Rev. D 23 2813;\\
         Kaminaga Y 1990 Class. Quantum Grav. 7 573.
\item  York I W 1985 Phys. Rev, D 28 2929.
\item  Balbinot R 1984 Class. Quantum Grav. 1 573; 1984 Phys. Lett. 136B 337; 
1986 Phys. Rev. D 33 1611
\item  Huang W-H Unpublished.
\item  Vaidya P C 1951 Proc. Ind. Acad. Sci. A 33 264.
\item  Bonnor W B and Vaidya P C 1970 Gen. Rel. Grav. 1 127.
\item  Davies P C W, Ford L H and Page D N 1986 Phys. Rev. D 34 1700.
\item  Balbinot Rand Bergamini R 1989 Phys. Rev. D 40 372.
\item  Mallet1 R L 1985 Phys. Rev. D 31 416: 1986 D 33 2201.
\item  Davies P C W 1977 Proc. R. Soc A 354 529.
\item  Gibbons G W and Hawking S W 1977 Phys. Rev. D 15 2738.
\item  Hawking S W and Ellis G F R 1973 "Large Scale Structure of Spacelime" 
(New York: Cambridge University Press)
\item  Lohiya D and Panchapakesan N 1978 J. Phys. A: Math. Gen. 11 1963;\\
         Mishima T and Nakayama A 1988 Phys. Rev. D 37 348; D 37 354.
\item  B. D. Koberlein and R. Mallet, ``Comment on ``Hawking Radiation of a Quantum Black Hole in an Inflationary Universe"", Class. Quantum Grav. 11 (1994) 283-285.
\item Raphael Bousso, Stephen Hawking, "(Anti-)Evaporation of Schwarzschild-de Sitter Black Holes"  Phys.Rev. D57 (1998) 2436-2442 [hep-th/9709224].\\
\item  Maulik K. Parikh and Frank Wilczek, "Global Structure of Evaporating Black Holes ", Phys.Lett. B449 (1999) 24-29 [gr-qc/9807031];  Maulik K. Parikh, "New Coordinates for de Sitter Space and de Sitter Radiation",  Phys.Lett. B546 (2002) 189-195 [hep-th/0204107]
\end{enumerate}
\end{document}